\documentclass[11pt,a4paper]{article}
\usepackage{amsmath,amssymb,amsthm,amsfonts,amstext,amsbsy,amscd}
\usepackage{epsfig}
\usepackage{color}

\newcommand{\R}{\mathbb{R}}

\newtheorem{thm}{Theorem}[section]
\newtheorem{prop}[thm]{Proposition}

\newtheorem{cor}[thm]{Corollary}

\begin{document}

\title{Modeling microstructure noise with mutually exciting point processes}

\author{E. Bacry\footnote{CMAP, Ecole Polytechnique, CNRS-UMR 7641, 91128 Palaiseau, France.},\, S. Delattre\footnote{Universit\'e Paris Diderot and CNRS-UMR 7599, 75013 Paris, France},\, M. Hoffmann\footnote{ENSAE-CREST and CNRS-UMR 8050, 92245 Malakoff Cedex, France.},\,\,J.F. Muzy\footnote{CNRS-UMR 6134, Universit\'e de Corse, 20250 Corte, France}$~^{,*}$.}
%\email{marc.hoffmann@ensae.fr}}

%\author{S. Delattre}
%\address{Universit\'e Paris Diderot and CNRS-UMR 7599, France}
%\email{delattre@math.jussieu.fr}

%\author{M. Hoffmann}
%\address{ENSAE-CREST and CNRS-UMR 8050, 3, avenue Pierre Larousse, 92245 Malakoff Cedex, France.}
%\email{marc.hoffmann@ensae.fr}

%\author{J.F. Muzy}
%\address{CNRS-UMR 6134, Universit\'e de Corse, 20250 Corte, France}
%\email{muzy@univ-corse.fr}

%\pacs{}

%\thanks{This research is part of the Chair {\it Financial Risks} of the {\it Risk
%Foundation}, the Chair {\it
%Derivatives of the Future} sponsored by the {F\'ed\'eration Bancaire
%Fran\c{c}aise} and the Chair {\it Finance and Sustainable Development}
%sponsored by EDF and Calyon.}

%\thanks{The financial data used in this paper have have been  provided by the company {\em QuantHouse
%EUROPE/ASIA}, http://www.quanthouse.com}.
\maketitle

\begin{abstract}
\textcolor{black}{We introduce a new stochastic model for the variations of asset prices at the
tick-by-tick level in dimension 1 (for a single asset) and 2 (for a pair of assets). The construction is based on marked point processes and relies on linear self and mutually exciting stochastic intensities as introduced by Hawkes. We associate a counting process with the positive and negative jumps of an asset price.  By coupling suitably the stochastic intensities of upward and downward changes of prices for several assets simultaneously, we can reproduce microstructure noise ({\it i.e.} strong microscopic mean reversion at the level of seconds to a few minutes) and the Epps effect ({\it i.e.} the decorrelation of the increments in microscopic scales) while preserving a standard Brownian diffusion behaviour on large scales.}
%\textcolor{black}{It enables to reproduce stylized microstructure facts such as the signature plot behaviour and the Epps effect on fine scales (at the level of seconds), while diffusing on large scales (at the level of hours and more).}

\textcolor{black}{More effectively, we obtain analytical closed-form formulae for the mean signature plot and the correlation of two price increments that enable to track across scales the effect of the mean-reversion up to the diffusive limit of the model. We show that the theoretical results are consistent with empirical fits on futures Euro-Bund and Euro-Bobl in several situations.}
\end{abstract}



%\noindent {\it e-mail:} emmanuel.bacry@polytechnique.fr, marc.hoffmann@ensae.fr, \\sylvain.delattre@math.jussieu.fr, muzy@univ-corse.fr \\

{\small \noindent This research is part of the Chair {\it Financial Risks} of the {\it Risk
Foundation}, the Chair {\it Derivatives of the Future} sponsored by the {F\'ed\'eration Bancaire
Fran\c{c}aise} and the Chair {\it Finance and Sustainable Development}
sponsored by EDF and Calyon.

\noindent The financial data used in this paper have been  provided by the company {\em QuantHouse
EUROPE/ASIA}, http://www.quanthouse.com.}

\newpage

\section{Introduction}
%\cite{Hewlett}
%\textcolor{black}{test}
With the availability of a huge number of quality high frequency data, there is a fast
growing literature devoted to the modelling of intra-daily asset prices behaviour.
Since Bachelier and the seminal work of Black and Scholes,
most popular models at a coarse time scale \textcolor{black}{-- for daily data, say -- are Brownian diffusions, see for instance the classical textbooks of Musiela \cite{Musiela} or Bouchaud and Potters \cite{BBoPo} and the references therein}. In particular, diffusion models aim at
describing more or less faithfully the volatility dynamics, characterized by stylized facts such as
volatility clustering or leverage effect \cite{BBoPo}. A key issue that naturally emerges
when one studies high frequency data is the problem of \textcolor{black}{improving} volatility estimation
or covariance estimation between two asset returns over a given time period, \textcolor{black}{thanks to the massive amount of data available at such scales nowadays}.
The discrete nature of time trade arrivals and of price variations
(prices are point processes living on a tick grid), the presence of
so-called microstructure noise (described as strong mean reversion effects at small scales) makes
this question highly non trivial. At a very high frequency, prices variations are also characterized by well documented
stylized facts like the signature plot behaviour and the Epps Effect \cite{Epps}.

\subsection{\textcolor{black}{High frequency volatility: microstructure noise}}

If $X(t)$ stands for the price of some asset at time $t$
(defined indifferently as the last traded price or the mid-price between best bid and best offer in the order book),
the signature plot can be defined
from the quadratic variation of $X(t)$ over a time periode $[0,T]$ at a scale $\tau >0$ \textcolor{black}{-- the so-called realized volatility -- as
\begin{equation}
\label{splot}
{\widehat C}(\tau) = \frac{1}{T}\sum_{n=0}^{T/\tau} \big|X\big((n+1)\tau\big)-X(n\tau)\big|^2.
\end{equation}}
The microstructure noise effect manifests
through an increase of the observed daily variance when one goes from large to small scales
{\it i.e.} in the limit $\tau \rightarrow 0$ (see {\it e.g.} Fig. \ref{figsplot1}(b)).
This behaviour is different from what one would expect
\textcolor{black}{if the data were sampled from a Brownian diffusion, in which case the function plotted in Fig. \ref{figsplot1}(b) should be flat. From the perspective of statistical estimation, this leads to a simple paradox. On one side, the smaller $\tau$, the larger is the dataset that can be use to estimate the volatility. However, how should one be using high-frequency data in order to obtain better estimates of the volatility, since the realized volatility \eqref{splot} is not stable as $\tau$ decreases.}

In the literature the most popular approaches attempts to model microstructure noise with the concept of
latent price. \textcolor{black}{One starts with a Brownian diffusion $X(t)$ defined as an efficient price, which is latent, in the sense that it cannot be observed directly. Instead the practitioner has only access to a
noisy version $\widetilde X(t)$ of $X(t)$ that accounts for microstructure noise.
The most successful version -- as far as mathematical development is concerned -- is the additive microstucture noise model, introduced in 2001 by Gloter and Jacod \cite{gloja1,gloja2} and in the context of financial data by Ait-Sahalia {\it et al.} \cite{AMZ, zha1, zha2}. Given a sampling scale $\tau$, one rather observes
\begin{equation} \label{def ctof}
\widetilde X(n\tau) = X(n\tau) + \xi_{n,\tau},
\end{equation}
where the microstructure noise term $\xi_{n,\tau}$ satisfies $\mathbb{E}\big[\xi_{n,\tau}\big]=0$ for obvious identifiability conditions. (Hereafter, $\mathbb{E}[\cdot]$ denotes the expectation operator.) The goal is then to separate the noise from the true signal $X(t)$, from which a classical volatility estimator can be performed. This  has raised a vast research program over the last decade, mostly covered by econometricians and statisticians, see \cite{AMZbis,seo, DS, RobRos, Ros, pod, BR1, BR2, barn, jac, MSH, mun, rei2, HMS} and the references therein. Whereas representation \eqref{def ctof} produces an elegant pilot model to describe microstructure noise effects at the scale of a few minutes, it cannot faithfully reproduce the data as they are observed on a microscopic scale of a few seconds: for instance, the discreteness of price changes is left out and the mathematical artefact of forcing $\widehat C(\tau)$ to explode when $\tau \rightarrow 0$ becomes unavoidable}.

\subsection{\textcolor{black}{High frequency correlations : the Epps effect}}

\textcolor{black}{Another important feature concerns the way different asset price movements are correlated. Stong correlations between asset returns at daily or larger time scales usually exist, and modern portfolio theory precisely relies on such dependences. One expects that coarse scale correlations originate from intraday strongly correlated movements, and this naturally raises the question of how to get better estimates of such correlations by using high frequency data. Again, an analogous paradox as for volatility estimation under microstructure noise arises.} In Fig. \ref{figcharts}, we display the last traded ask prices of futures Bund 10Y and Bobl 5Y over a period of a few hours, prices increments
are clearly correlated in their intra-day variations. If $X_1(t)$ and $X_2(t)$ are
the prices of two assets, a correlation coefficient estimator over a time period $[0,T]$
can be naturally defined from high frequency price increments as\textcolor{black}{
$$\rho(\tau) = \frac{{\widehat C_{12}}({\tau})}{{\sqrt{\widehat C_1(\tau) \widehat C_2(\tau)}}},$$
where
\begin{equation}
\nonumber
\widehat C_{12}({\tau}) =  \frac{1}{T}\sum_{n=0}^{T/\tau} \left[X_1\big((n+1)\tau\big)-X_1\big(n\tau\big)\right]\left[X_2\big((n+1)\tau\big)-X_2\big(n\tau\big)\right]
\end{equation}
and $\widehat C_1(\tau)$ and $\widehat C_2(\tau)$ denote the realized volatility of $X_1(t)$ and $X_2(t)$ respectively at scale $\tau$.}
In Fig. \ref{fig2D}(c), we plot $\rho(\tau)$ as a function of $\tau$ for
the couple Bund/Bobl. It is striking to observe that the correlation coefficient is an increasing
function of the time resolution and that correlation almost vanish at a very high frequency. This
phenomenon, first reported by Epps \cite{Epps}, is the so-called Epps effect. Few approaches however address the Epps effect in the literature. \\

Besides the problems related to the estimation of coarse scale asset properties,
we see that modelling high frequency price dynamics is a source a many challenging questions. Other approaches consist in defining  ``fine" scale models and addressing directly the price dynamics at the tick level, see for instance Engle and collaborators, \cite{EnRu98,EnLu01} \textcolor{black}{who introduce the ACM-ACD model or Bauwens and Hautsch \cite{Hautsch}, Bowsher \cite{Bowsher} who construct intensity based point processes for modelling high frequency data. In this context, Hewlett \cite{Hewlett} introduces a model for measuring trade arrivals that is formally very close to our construction. However, Hewlett's approach is mostly order book oriented: it focuses on the imbalance properties of an asset and leaves out the directional behaviour of the prices. In the same way, the aforementioned  literature that uses point processes is essentially specialized to high frequency data description, or focuses on trades arrivals dynamics. In particular, the questions related to intermediate or asymptotic behaviour of the statistical price properties as the resolution scale varies are not considered. We plan to address in the present paper this next logical step.}

\subsection{\textcolor{black}{Objective and content of the paper}}

We introduce a ``fine-to-coarse" model that \textcolor{black}{starts from the description of the changes of prices in continuous time} and that allows one, from the microscopic properties of the model to recover a large scale diffusion behaviour.
More precisely, be defining a (multivariate) tick-by-tick model by means \textcolor{black}{of marked point processes with appropriate stochastic intensities}, we are able to control its features at all scale ({\it i.e.} its aggregation properties).
%and notably we can account for the signature plot behavior and the Epps effect simultanesouly.
Our model relies on multivariate Hawkes
processes \cite{Hawkes, DVJ} associated with positive and negative jumps of the asset prices. Notably, \textcolor{black}{by coupling suitably the stochastic intensities of upward and downward changes of prices for several assets simultaneously, we can reproduce microstructure noise ({\it i.e.} strong microscopic mean reversion) and the Epps effect ({\it i.e.} the decorrelation of the increments in microscopic scales) while preserving a standard Brownian diffusion behaviour on large scales.} \\

The paper is organized as follows: in Section \ref{1dmodel} we define a univariate version of the
model and compute the expected signature plot. We discuss how \textcolor{black}{to simulate the model in practice and how to estimate its parameters from real data}. The vector (bivariate) version of the model is defined in Section \ref{2dmodel}.
We show that a closed form expression for the price variations correlation functions
can be explicitly obtained for all time scales and time lags. From such quantity one can deduce all second order properties of the process like individual signature plots, the Epps effect or lead-lag effects. We discuss
the large scale diffusive limit of our model as far as its correlations properties
are concerned and make the link with \textcolor{black}{further mathematical asymptotic results (that are presented in a separate forthcoming paper).} Comparisons to empirical data are provided
in Section \ref{data} in both 1D and 2D frameworks. Section \ref{conclusion} is devoted to a conclusion and
prospects for future research.

\section{\textcolor{black}{The model in the univariate case}}
\label{1dmodel}

\subsection{Construction of the model} \label{construction 1D}
We start with two point processes $N_1(t)$ and $N_2(t)$ \textcolor{black}{for $t \in [0,T]$} that represent
respectively the sum of positive and negative jumps of some asset price $X(t)$ \textcolor{black}{over some time horizon $[0,T]$}:
\[
  X(t) = N_1(t)-N_2(t).
\]
If $N_1(t)$ and $N_2(t)$ are two independent Poisson processes with intensity $\mu$, it is easy
to show that the model diffuses at large scale, \textcolor{black}{{\it i.e.}, when $T \rightarrow \infty$, by introducing the scaling factor $T^{1/2}$, we obtain the following limit in distribution:}
$$\lim_{T \rightarrow +\infty} \frac{1}{\sqrt{T}} X(tT)  \stackrel{(d)}{=}\sqrt{2\mu} B(t),\;\;t \in [0,1],$$
where $\sqrt{2\mu}$ is the diffusive \textcolor{black}{or macroscopic volatility, that accounts for the activity of negative and positive jumps, hence the factor $2$ in the limit}. According to Eq. \eqref{splot},
the corresponding mean signature plot is flat: for all $\tau >0$, we have $\mathbb{E}[\widehat C(\tau)] = 2\mu$.
In order to account for the previously reported noise microstructure features,
intuitively and as confirmed by empirical observations,
one has to introduce some mean reversion \textcolor{black}{in the small scales, while ensuring that this mean reversion effect vanishes on large scale}. This can be naturally done within the context of (multivariate)
Hawkes process \cite{Hawkes, DVJ} as follows.\\

\textcolor{black}{For technical reason, we extend in a first step the time horizon $[0,T]$ over the whole real line $\R=(-\infty,+\infty)$}. Let $\lambda_i(t)$ \textcolor{black}{ for $t \in \R$} be the stochastic intensities of \textcolor{black}{two counting processes} $N_i(t)$, $i=1,2$, \textcolor{black}{such that} at time $t$:
\begin{equation}
   \lambda_{i}(t) = \lim_{\Delta  \rightarrow 0} \Delta^{-1} \mathbb{E} \left[ N_i(t+\Delta)-N_i(t)\; | \;{\mathcal F}_t \right]
\end{equation}
where ${\mathcal F}_t$ stands for the filtration generated by the history of the processes $N_1(t)$, $N_2(t)$. The bivariate process $\big\{N_1(t),N_2(t)\big\}$ is a \textcolor{black}{linear Hawkes process if $N_1(t)$ and $N_2(t)$ have no common jumps and if there exist}
four nonnegative functions $\{\varphi_{ij}\}_{i,j=1,2}$ such that
\begin{equation}
   \lambda_{i}(t) = \mu_{i} + \int_{-\infty}^{t} \varphi_{ii}(t-u) dN_i(u) + \int_{-\infty}^{t} \varphi_{ij}(t-u) dN_j(u).
\end{equation}
The so-obtained process can be shown to be well defined and \textcolor{black}{to admit a version} with stationary increments under the stability condition\begin{equation}
\label{stability}
\mbox{all the eigenvalues of  the matrix } \{||\varphi_{ij}||_1\} \mbox{are} < 1,
\end{equation}
\textcolor{black}{where $\|\varphi\|_1 = \int_{\R}\varphi(t)dt$},
see \cite{Hawkes, DVJ}. Mean reversion can be translated by the fact that the more $X(t)$ goes up, the greater the intensity $\lambda_2(t)$
and conversely, the more $X(t)$ goes down, the greater the intensity $\lambda_1(t)$.
This leads to the following simplified version of previous model (where only mean-reverting terms were kept):
\begin{eqnarray}
\label{defl1}
\lambda_{1} (t)  & = &\mu +\int_{-\infty}^t \varphi(t-s)dN_{2}(s) \\
\label{defl2}
\lambda_{2} (t)  & = & \mu +\int_{-\infty}^t \varphi(t-s)dN_{1}(s)
\end{eqnarray}
where $\mu$ is an exogenous intensity and
$\varphi(t)$ a positive kernel which is causal (i.e., Supp$(\varphi) \subset \R^+$).
Eqs. \eqref{defl1} and \eqref{defl2} define two mutually exciting point processes that
are stationary and stable under the condition  $||\varphi||_1 < 1$.

A simple and natural choice for $\varphi$ is a right-sided exponential function:
\begin{equation}
\label{expokernel}
\varphi(t) = \alpha e^{-\beta t} 1_{\R^+}(t)
\end{equation}
where $\alpha,\beta >0$ are such that
\begin{equation} \label{stab condit}
||\varphi||_1 = \frac{\alpha}{\beta}<1.
\end{equation}

\subsection{Signature plot}
%If we chose the initial condition $X(0) = 0$\footnote{Let us note that if we impose such an initial condition, strictly speaking, the process does not have stationary increments. The increments are only asymptotically stationary. In the following, we will implicitely consider that we are in the  asymptotic limit (i.e., time goes to infinity). Consequently we can consider that the increments {\em are} stationary)} then
\textcolor{black}{We are interested only in $X(t)$ on $[0,T] \subset \R_+$ and for simplicity and without loss of generality, we set\footnote{\textcolor{black}{Indeed, since only the increments of the processes $N_i(t)$ come into play, we may (and will) assume that $N_1(t)=N_2(t)=0$ hence $X(0)=0$.}} $X(0)=0$.} Using Eq. \eqref{splot}, the mean signature plot can be written as
 \begin{equation}
\nonumber
 C(\tau) = \mathbb{E}\big[{\widehat C}(\tau)\big] =  \frac{1}{\tau} \mathbb{E}\big[ |X((n+1)\tau)-X(n\tau)|^2\big] = \frac 1 {\tau} \mathbb{E}[X(\tau)^2].
\end{equation}
If $\varphi(t)$ is defined as in Eq. \eqref{expokernel}, then one can actually obtain  a closed form for the mean signature plot using the approach initiated in \cite{Bartlett,Hawkes}.
In Appendix 1, we prove the following result:
\begin{prop} Under the stability condition \eqref{stab condit}, we have
\label{prop1}
\begin{equation}
\label{1Dsplot}
 C(\tau)  = \Lambda \left({\kappa^2} + (1-{\kappa^2})\frac{1-e^{-\gamma \tau}}{\gamma \tau}\right)
 \end{equation}
where
$$\Lambda = \frac{2\mu}{1-||\varphi||_1},\;\;\kappa = \frac 1 {1+||\varphi||_1},\;\;\text{and}\;\;\gamma = \alpha+\beta.$$
\end{prop}

One sees in particular a cross-over from the microstructural variance
$$\textcolor{black}{V_0 = \mathbb{E}[\widehat C(0)] = \Lambda =  2\mathbb{E}(\lambda_i)},$$
to the diffusive variance
$$V_{\infty} = \mathbb{E}[\widehat C(\infty)] =  \Lambda\kappa^2,$$
In  Fig. 1(b), we display a plot example of the function $C(\tau)$.

\subsection{Numerical simulations and parameter estimation}
\label{est1d}
In this section, we focus on the numerical simulations and the parameters estimation of the simplified univariate model defined by Eqs (\ref{defl1}) and (\ref{defl2}) in the case the function $\varphi$ is the right sided exponential function (\ref{expokernel}). Thus, there are 3 parameters, namely : $\theta = (\mu,\alpha,\beta)$.

Simulation of this process on an interval $[0,T]$ can be performed using the thinning algorithm described in \cite{Ogata81}. It basically consists in simulating on $[0,T]$ a standard Poisson process with an intensity $M$ large enough
such that it satisfies the following condition:
\textcolor{black}{
$$\lambda_1(t) < M\;\;\text{and}\;\;\lambda_2(t) < M,\;\;\text{for all}\;\; t \in [0,T]$$
(which of course can be checked {\em only} {\it a posteriori}). A thinning procedure is then applied to each jump of the obtained process from the first one to the last one allowing to either reject the point (with probability $M-\lambda_1(t)-\lambda_2(t)$) or mark it as a jump of $N_1(t)$ with probability $\lambda_1(t)$ or of $N_2(t)$ with probability $\lambda_2(t)$, where $t$ is the time of the considered jump.}\\

\begin{figure}
\begin{center}
\includegraphics[width=10cm]{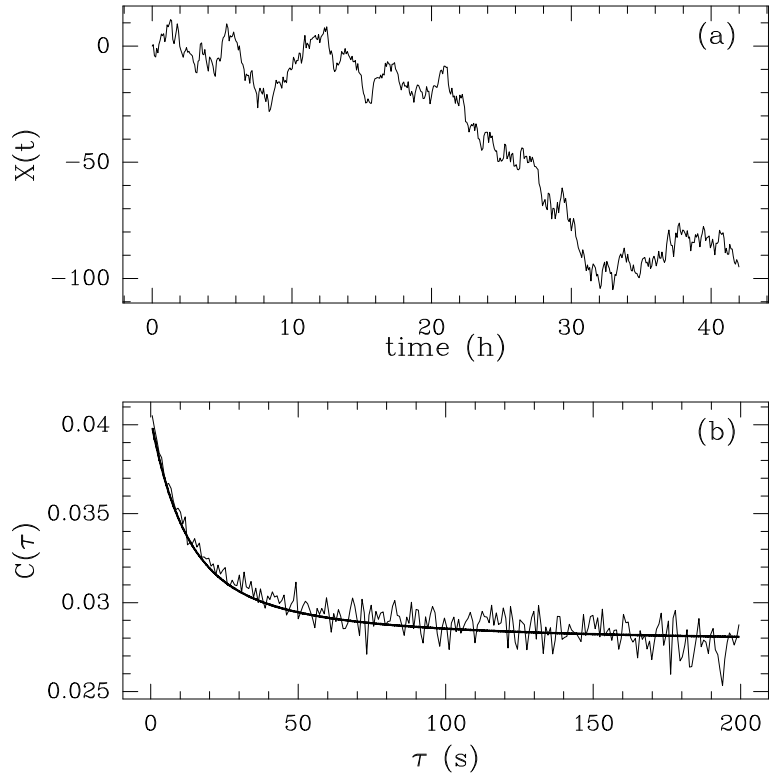}
\end{center}
\caption{
\label{figsimu}
\textcolor{black}{Numerical simulation of a 1D price model using mutually exciting Hawkes processes with parameters $\mu = 0.16$, $\alpha = 0.024$, $\beta = 0.11$ (Let us note that these values were chosen to match the estimated parameters on real data as shown in Section \ref{realest})}.
(a) Simulated sample path $X(t)$ (42 hours long) (b) Estimated signature plot $C(\tau)$
with the corresponding
expected analytical shape according to Eq. \eqref{1Dsplot}. $\tau$ is expressed in seconds.
}
\end{figure}

A realization of the process $X$ over $T=42$ hours is represented in Fig. \ref{figsimu}(a)
with $\theta = (\mu = 0.016, \alpha = 0.023,\beta = 0.11)$. Let us note that $\mu$,$\alpha$ and $\beta$ are all expressed in the  the same unit, namely $\text{seconds}^{-1}$.
These particular values were chosen to match the estimated parameters on real data (see Section \ref{realest}).

Let $\{t^{(1)}_{i}\}_{0\le i<N_1(\cdot)}$
(resp. $\{t^{(2)}_{i}\}_{0\le i<N_2(\cdot)}$) the upward (resp. downward) jumps of the realization of $X$. The estimation of the parameters can be processed in very different ways depending on what is the focus of the model.
On the one hand, if one is mainly interested \textcolor{black}{in the ability of the model to reproduce} the  mean signature plot, the parameters can be estimated using a best fit of the realized signature plot. \textcolor{black}{The realized signature plot over $[0,T]$ is defined as
\begin{equation}
\label{est_reg}
\widehat C(\tau) = \frac{1}{T}\sum_{n=0}^{T/\tau} \big| X\big((n+1)\tau\big)- X(n\tau)\big|^2,
\end{equation}}
and {\em regression} estimator $\widehat \theta_{\text{reg}}$ is then naturally given by
\begin{equation}
\widehat \theta_{\text{reg}} = \text{Argmin}_{\theta} \big|\widehat C(\tau) -  C(\tau)\big|^2,
\end{equation}
where $C(\tau) $ is defined by (\ref{1Dsplot}).

On the other hand, if the goal of the model is not simply to reproduce the signature plot behaviour but to mimick the arrival times themselves, it is more natural to consider the Maximum Likelihood Estimator (MLE) instead. This is possible since
there is a closed formula for the likelihood. \textcolor{black}{Let $N(t)$ be a multivariate point process with conditionnal intensity $\lambda(t) = \lambda_\theta(t)$ depending on a parameter $\theta$. If, for every $\theta$, the law of the process $N(t)$ is absolutely continuous w.r.t. to the law of a standard stationary Poisson process, then the statistical model generated by the continuous observation of $N(t)$ over $[0,T]$ has a log-likelihood function given by
\begin{equation} \label{est_mle}
L(\theta) = \int_0^T \log\lambda_\theta(t)dN(t) + \int_0^T\big(1-\lambda_\theta(t)\big)dt,
\end{equation}
see for instance \cite{JS, DVJ} or \cite{Ogata78} in the context of Hawkes processes and the references therein.
Consequently the likelihood functions $\theta \mapsto L_1(\theta)$ and $\theta \mapsto L_2(\theta)$ generated by the observation of $N_1(t)$ and  $N_2(t)$ over $[0,T]$ are given by
\begin{align*}
L_{1}(\theta)  = &\sum_{0\le t_i^{(1)}< N_1(T)} \log \Big( \mu + \sum_{0\le t_j^{(2)}<N_2(T)} \alpha e^{-\beta(t_i^{(1)}-t_j^{(2)})} \Big) \\
& - (\mu-1) T - \sum_{0\le t_j^{(2)} < N_2(T)} \frac \alpha \beta \left(1-e^{-\beta(T-t_j^{(2)})} \right).
\end{align*}
and similarly
\begin{align*}
L_{2}(\theta) = & \sum_{0\le t_i^{(2)}< N_2(T)} \log \Big( \mu + \sum_{0\le t_j^{(1)}<N_2(T)} \alpha e^{-\beta(t_i^{(2)}-t_j^{(1)})} \Big) \\
&- (\mu-1) T - \sum_{0\le t_j^{(1)} < N_1(T)} \frac \alpha \beta \left(1-e^{-\beta(T-t_j^{(1)})} \right).
\end{align*}
and the MLE of $\theta$ based on the observation of $X(t)$ over $[0,T]$ is thus given by}
\begin{equation}
\widehat \theta_{\text{MLE}} = \text{Argmin}_\theta \big(L_1(\theta)+L_2(\theta)\big).
\end{equation}

Of course, for both estimators $\widehat \theta_{\text{reg}}$ and $\widehat \theta_{MLE}$ the minimization (Eqs (\ref{est_reg}) and (\ref{est_mle})) must be performed under the constraints
\begin{equation}
\mu > 0,~~\alpha > 0,~~\beta > 0,~~{\mbox{and the stability condition :}}~\frac \alpha \beta < 1.
\end{equation}
Let us note that whereas the regression estimator $\widehat \theta_{\text{reg}}$ can be performed on uniformly sampled data, the ML estimator $\widehat\theta_{MLE}$ needs to have access to the point process itself. In that sense, $\widehat \theta_{reg}$ can be considered as a (multi-scale) low-frequency estimator. For that reason, when applied to real data, we expect it to be much more stable than $\widehat \theta_{MLE}$ (see Section \ref{data}). Moreover, the regression estimator has the advantage to be computationally faster than the MLE.

The estimated parameters of the realization shown in Fig. \ref{figsimu}(a) using MLE are
$\widehat \mu = 0.016$, $\widehat \alpha = 0.024$ and $\widehat \beta = 0.11$. They match the true parameter values ($\mu = 0.16$, $\alpha = 0.024$, $\beta = 0.11$).

The estimation using the regression estimator appears to be almost as accurate as the MLE: we obtain $\widehat \mu = 0.016$, $\widehat \alpha = 0.021$ and $\widehat \beta = 0.010$.
In Fig. \ref{figsimu}(b) we display
the estimated and theoretical signature plots (Eq. \eqref{1Dsplot}). One sees that
the latter provides a good fit of the data.

\section{\textcolor{black}{The model in the bivariate case}}
\label{2dmodel}

\subsection{Definition}
The bivariate model is a natural extension of the former construction.
We start from two processes $X_1(t)$ and $X_2(t)$ \textcolor{black}{with $t \in \R$}, each constructed as in Section \ref{construction 1D} ({\it i.e.} in dimension 1) and introduce a supplementary coupling on the intensities of the processes in order to create a dependence structure.
More precisely, we consider four mutually exciting point processes  $\{N_i(t)\}_{i=1,\ldots,4}$
associated with the positive and negative variations of $X_1(t)$ and $X_2(t)$:
\begin{eqnarray*}
   X_1(t) & = & N_1(t)-N_2(t) \\
   X_2(t) & = & N_3(t)-N_4(t) \; .
\end{eqnarray*}
The \textcolor{black}{joint law of the processes $N_i(t)$} is characterized by their intensities:
\begin{eqnarray*}
   \lambda_{i}(t) & = & \mu_{i} + \sum_{j=1}^4 \int_{-\infty}^t \varphi_{ij}(t-s) \; dN_j(s) \; \; i=1,\ldots,4
\end{eqnarray*}
Let us note that, as previously, the kernels $\varphi_{ij}$ account for the mutual and cross
excitations of positive/negative parts of the couple of assets.
In the sequel, \textcolor{black}{in accordance with the univariate case},  we suppose that $\mu_1 = \mu_2$ and $\mu_3 = \mu_4$ \textcolor{black}{(but we do not assume that $\mu_1=\mu_3$)}.

In order to account for mean reversion and cross coupling between the two assets, we
do not consider all possible mutual and cross excitations. More precisely, we only
want to consider Upward-$X_1$-Upward-$X_2$ and Downward-$X_1$-Downward-$X_2$ couplings.
We ignore further possible couplings Upward-$X_1$-Downward-$X_2$ and Downward-$X_1$-Upward-$X_2$
between $X_1$ and $X_2$.
We thus choose the matrix $\mathbf{\Phi} =\{\varphi_{ij}\}_{1 \leq i,j \leq 4}$ of the following form
\begin{equation}
\label{phimatrix}
{\mathbf \Phi} = \left(
\begin{array}{cccc}
0 & \varphi_{12} & \varphi_{13} & 0 \\
\varphi_{12} & 0 & 0 & \varphi_{13} \\
\varphi_{31} & 0 & 0 & \varphi_{34} \\
0 & \varphi_{31} & \varphi_{34} & 0
\end{array}
\right).
\end{equation}

Let us consider the matrix ${\bf \Gamma} = \{\Gamma_{ij}\}_{1 \leq i,j \leq 4}$ with entries
$\Gamma_{ij} = ||\varphi_{ij}||_1$ and ${\bf \Lambda}$ the vector of mean intensities:
\begin{equation}
\label{deflambda}
 \Lambda_{i}  =  \mathbb{E}\left[\frac{dN_{i}}{dt} \right]
\end{equation}
It is shown in Hawkes \cite{Hawkes} that:
\begin{equation}
\label{lambda}
          {\bf \Lambda} = ({\bf Id -\Gamma})^{-1} {{\mathbf \mu}}
\end{equation}
The solution to this equation reads:
\begin{equation}
\label{lambda}
{\mathbf \Lambda} =
\left[ \begin {array}{c} -{\frac {{ \mu_1}\,{ \Gamma_{34}}-{ \mu_1}-{ \Gamma_{13}}\,{ \mu_3}}{{ \Gamma_{12}}\,{ \Gamma_{34}}-
{ \Gamma_{12}}-{ \Gamma_{34}}+1-{ \Gamma_{31}}\,{ \Gamma_{13}}}}
\\\noalign{\medskip}-{\frac {{ \mu_1}\,{ \Gamma_{34}}-{ \mu_1}-
{ \Gamma_{13}}\,{ \mu_3}}{{ \Gamma_{12}}\,{ \Gamma_{34}}-{
\Gamma_{12}}-{ \Gamma_{34}}+1-{ \Gamma_{31}}\,{ \Gamma_{13}}}}
\\\noalign{\medskip}{\frac {-{ \mu_3}\,{ \Gamma_{12}}+{
\Gamma_{31}}\,{ \mu_1}+{ \mu_3}}{{ \Gamma_{12}}\,{ \Gamma_{34}}-
{ \Gamma_{12}}-{ \Gamma_{34}}+1-{ \Gamma_{31}}\,{ \Gamma_{13}}}}
\\\noalign{\medskip}{\frac {-{ \mu_3}\,{ \Gamma_{12}}+{
\Gamma_{31}}\,{ \mu_1}+{ \mu_3}}{{ \Gamma_{12}}\,{ \Gamma_{34}}-
{ \Gamma_{12}}-{ \Gamma_{34}}+1-{ \Gamma_{31}}\,{ \Gamma_{13}}}}
\end {array} \right]
\end{equation}

As for the univariate case, we consider functions $\varphi_{ij}$ that are causal exponentials:
\begin{equation}
\label{phiij}
  \varphi_{ij}(x) = \alpha_{ij} e^{-\beta_{ij} x}{\bf 1}_{\R_+}(x).
\end{equation}
In that case one simply obtains:
\begin{equation}
 \Gamma_{ij} = \frac{\alpha_{ij}}{\beta_{ij}}.
\end{equation}

\textcolor{black}{For some $t_0 \in \R$ and a scale $\tau$, let us define the increments of a process $X$ as
$$\Delta_\tau X(t_0) = X(t_0+\tau)-X(t_0).$$}
If one wants to characterize multiscale self and cross-correlations of two asset prices  $X_1(t)$ and $X_2(t)$
the quantity of interest is the
covariance matrix ${\bf C}(t,\tau) = \{C_{kl}(t,\tau)\}_{1 \leq k,l \leq 2}$ with entries
\begin{equation}
\label{cmatrix}
    C_{kl}(\tau,t)  = \mbox{Cov}\left[\Delta_\tau X_k (t_0),\Delta_\tau X_l (t_0+t)\right].
\end{equation}
Thanks to the stationarity of the increments of $X_k$, the matrix $\mathbf{C}$ does not depend on $t_0$.
\textcolor{black}{In particular, we recover from ${\bf C}(t,\tau)$ the mean signature plots on the diagonal
and the Epps effect off diagonal as $\tau$ varies. We also can estimate possible lead-lag effects across various time scales, that we can loosely define as the property that the function $t \mapsto C_{kl}(\tau,t)$ is not symmetric around $0$ for $k \neq l$ (see below).}
In the following, we denote by $C_{kl}(\tau) = C_{kl}(\tau,t=0)$ the correlations at lag $t=0$.

\subsection{Computation of the signature plot and the Epps effect}

In Appendix 2, we show that we obtain a closed form expression for
the covariance matrix \eqref{cmatrix}.
The general expression where all parameters are arbitrary is cumbersome
and we limit ourselves to a matrix  $\mathbf{\Phi}$ of the form \eqref{phimatrix} with \textcolor{black}{$\varphi_{ij}(x)=\alpha_{ij}\exp(-\beta_{ij}x){\bf 1}_{\R_+}(x)$} and
$$\beta_{ij} = \beta,\;\;i,j=1,3.$$
In that case we get the following proposition:
\begin{prop}
\label{prop2}
The expression of the covariance $C_{kl}(\tau)$ as a function of the time scale $\tau$ reads:
{\footnotesize
\begin{eqnarray*}
\label{e1}
&\displaystyle \frac{C_{11}(\tau)}{\tau} & =  \frac{2 A_{11}e^{-G_1\tau}}{G_1^2 \tau}+
\frac{2 B_{11}e^{-G_2\tau}}{G_2^2\tau}+2 \Lambda_1
+\frac{2 A_{11}}{G_1}+\frac{2 B_{11}}{G_2}
-\frac{2 A_{11}}{G_1^2\tau}-\frac{2 B_{11}}{G_2^2\tau}, \\
\label{e2}
&\displaystyle  \frac{C_{21}(\tau)}{\tau} &=  \frac{C_{12}(\tau)}{\tau} =  (A_{21}+A_{12})(\frac{e^{-G_1\tau}-1}{G_1^2 \tau}
+\frac{1}{G_1})
+(B_{21}+B_{12})(\frac{e^{-G_2\tau}-1}{G_2^2\tau}+\frac{1}{G_2}),\\
\label{e3}
&\displaystyle  \frac{C_{22}(\tau)}{\tau} & =  \frac{2 A_{22}e^{-G_1\tau}}{G_1^2 \tau}+
\frac{2 B_{22}e^{-G_2\tau}}{G_2^2\tau}+2 \Lambda_3
+\frac{2A_{22}}{G_1}+\frac{2B_{22}}{G_2}
-\frac{2 A_{22}}{G_1^2\tau}-\frac{2 B_{22}}{G_2^2\tau},
\end{eqnarray*}
}
\textcolor{black}{with explicit} constants provided in Appendix 2.
\end{prop}

It is interesting to discuss the fully symmetric case, {\it i.e.}
when $\varphi_{12} = \varphi_{34}$, $\varphi_{13} = \varphi_{31}$ and
$\mu_1 = \mu_3 = \mu$.
In this case a direct computation from \eqref{lambda} leads to
\begin{equation}
   \Lambda = \Lambda_1 = \Lambda_2 = \Lambda_3 = \Lambda_4 = \frac{\mu}{1-\Gamma_{12}-\Gamma_{13}}
\end{equation}

If follows from the results of Appendix 2, that
\begin{eqnarray*}
   Q_1 &= Q_4 = & \frac{-\mu \left( \Gamma_{12}^2+\Gamma_{12}-\Gamma_{13}^2\right)}{\left((\Gamma_{12}+1)^2-\Gamma_{13}^2\right)\left(1-\Gamma_{12}-\Gamma_{13}\right)} \\
   Q_2 & = Q_3 = & \frac{-\mu \Gamma_{13}}{\left((\Gamma_{12}+1)^2-\Gamma_{13}^2\right)\left(1-\Gamma_{12}-\Gamma_{13}\right)}
\end{eqnarray*}

After some algebra, one obtains:
\begin{cor}
In the fully symmetric case, the covariance matrix has the following expression:
\label{cor1}
\begin{align*}
   \frac{C_{11}(\tau)}{\tau} = & \Lambda+\frac{RC_1}{2 G_1}+\frac{RC_2}{2 G_2} \\
& +R{\frac {{ C_2}{{G_1}}^{2}{{\rm e}^{-{
\tau}\,{G_2}}}-{ C_1}{{G_2}}^{2}+{ Q_1}{{G_2}}^{2}{
{\rm e}^{-{\tau}\,{G_1}}}-{ C_2}{{G_1}}^{2}}{{{ 2 G_2}}^
{2}{{G_1}}^{2}{\tau}}},
\end{align*}
and
\begin{align*}
\frac{C_{12}(\tau)}{\tau}  = & \frac{-RC_1}{2 G_1}+\frac{RC_2}{2 G_2} \\
&+ {\frac {R \left( { C_1}{{ \lambda_2}}^{2}-{ C_2}{{G_1}}^{2}-{
 C_1}{{G_2}}^{2}{{\rm e}^{-{ \lambda_1}{\tau}}}+{ C_2}\,{{
 G_1}}^{2}{{\rm e}^{-{G_2}{\tau}}} \right) }{{{ 2 G_2}}^{2}{
{G_1}}^{2}{\tau}}}
\end{align*}
where
\begin{eqnarray*}
   R & = & \frac{\beta \mu}{\Gamma_{12}+\Gamma_{13}-1} \\
   C_1 & = & \frac{(2+\Gamma_{12}+\Gamma_{13})(\Gamma_{12}+\Gamma_{13})}{1+\Gamma_{12}+\Gamma_{13}} \\
   C_2 & = & \frac{(2+\Gamma_{12}-\Gamma_{13})(\Gamma_{12}-\Gamma_{13})}{1+\Gamma_{12}-\Gamma_{13}}
 \end{eqnarray*}
 and
$$
  G_1   =  \beta(1+\Gamma_{12}+\Gamma_{13}),\;\;
  G_2   =  \beta(1+\Gamma_{12}-\Gamma_{13}).$$
\end{cor}

One can reproduce the Epps effect by evaluating the behavior
of the correlation coefficient $\rho(\tau) = C_{12}(\tau)/C_{11}(\tau)$ as a function of $\tau$.
From Corollary (\ref{cor1}) one can see that
\begin{eqnarray*}
&  \rho(\tau) & = \frac{R(Q_2-Q_1)}{4 \Lambda} \tau +O\left( \tau^2 \right)\;\;\text{as}\;\;\tau \rightarrow 0, \\
&  \rho(\tau) & \rightarrow  \frac{R_2-R_1}{2\Lambda+R_1+R_2}\;\;\text{as}\;\;\tau \rightarrow \infty,
\end{eqnarray*}
where we have set $R_i = R Q_i/\lambda_i$.
If one considers the definition of each constant, one gets
\begin{equation}
\label{correl}
\rho(\tau) \rightarrow \frac{2 \Gamma_{13}(1+\Gamma_{12})}{1+\Gamma_{13}^2+2\Gamma_{12}+\Gamma_{12}^2}\;\;\text{as}\;\;\tau \rightarrow \infty.
\end{equation}

Notice that when the processes are not correlated, {\it i.e.} when $\Gamma_{13} = 0$, we have
$G_1 = G_2 = \beta(1+\Gamma_{12}) = \beta+\alpha_{12} = \gamma$, $\Lambda = \mu/(1-\Gamma_{12})$,
$R = \beta\mu / (\Gamma_{12}-1)$, $Q = Q_1 = Q_2 = \Gamma_{12}(\Gamma_{12}+2)/(1+\Gamma_{12})$. Thus
$$\frac{RQ}{\lambda_1} = \mu \frac{\Gamma_{12}(\Gamma_{12}+2)}{(\Gamma_{12}-1)(\Gamma_{12}+1)^2}$$
and therefore
\begin{eqnarray*}
  \frac{C_{11}(\tau)}{\tau} & = & \Lambda+\frac{RQ}{G_1}+\frac{RQ}{G_1^2 \tau}(e^{-G_1 \tau}-1) \\
                           & = & \frac{\mu}{1-\Gamma_{12}}\left(\frac{1}{(1+\Gamma_{12})^2}+(1-\frac{1}{(1+\Gamma_{12})^2})\frac{1-e^{-\gamma \tau}}{\gamma \tau}\right).
\end{eqnarray*}
We thus recover in that case the univariate result of Eq. \eqref{figeppssimu}.

In order to account for the existence of a possible lead-lag effect, one can introduce
a measure of the asymmetry of the covariance matrix at scale $\tau$ as e.g.
\begin{equation}
  \Delta(\tau) = C_{12}(\tau,\tau)-C_{21}(\tau,\tau) \; .
\end{equation}
This coefficient measures the difference of the correlation of assets 1 and 2 returns
at scale $\tau$ and lag $\tau$ when assets 1 is respectively in the past or in the future of asset 2. It is easy to see that if the matrix \eqref{phimatrix} is symmetric then $\Delta = 0$ and there is no lead-lag. In the general case, from the results (and within notations)
of Appendix 2, if $D_1 = A_{12}-A_{21}$ and $D_2 = B_{12}-B_{21}$ then:
\begin{equation}
  \Delta(\tau) = D_1 \frac{1+e^{-2 G_1 \tau}2-2 e^{-G_1 \tau}}{G_1^2} +  D_2 \frac{1+e^{-2 G_2 \tau}2-2 e^{-G_2 \tau}}{G_2^2}
\end{equation}
The expressions for $D_1$ and $D_2$ are relatively heavy to handle in the general
case. In the simple asymmetric case when $\alpha_{13} = 0$ one has:
\begin{equation}
D_1  = -\frac{2 \alpha_{12} \alpha_{31} \beta \mu_1 (2\beta+\alpha_{12})}{(\alpha_{12}-\alpha_{34})(2 \beta+\alpha_{34}+\alpha_{12}) (\beta-\alpha_{12})}
\end{equation}
The specific study of lead-lags effects within this approach will be the scope of a forthcoming
paper.

\subsection{Numerical simulations and parameter estimation}
\label{est2d}
In this section, we focus on numerical simulations and parameters estimation of the bivariate model in the fully symmetric case ($\varphi_{12}=\varphi_{34}$,  $\varphi_{13}=\varphi_{31}$ and $\mu_1=\mu_3=\mu$),
 in the case where all the functions $\varphi_{ij}$ are right sided exponential functions (\ref{phiij}) with $\beta_{ij}=\beta$ for all $i,j$. Thus, there are 4 parameters, namely : $\theta = (\mu,\alpha_{1,2},\alpha_{1,3},\beta)$.

\begin{figure}[h]
\begin{center}
\includegraphics[width=10cm]{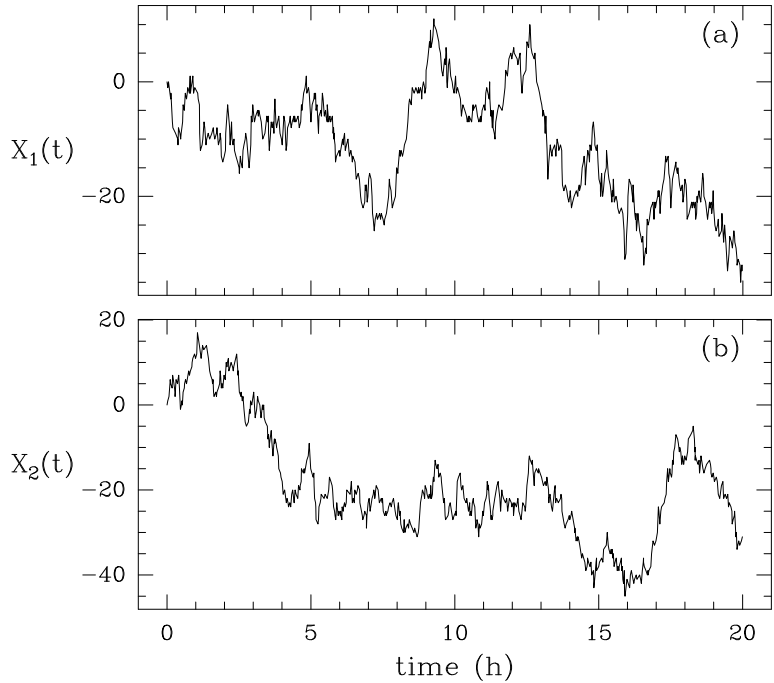}
\end{center}
\caption{
\label{figc1}
Numerical simulation of a 2D symmetric price model using Hawkes processes with parameters
$\alpha_{12} = 0.23$, $\alpha_{13} = 0.01$, $\beta = 0.11$ and $\mu = 0.015$.
(a) Sample path of $X_1(t)$  (b) Sample path of $X_2(t)$. The asymptotic correlation
between large scale increments is $\rho = 0.15$.
}
\end{figure}

\begin{figure}[h]
\begin{center}
\includegraphics[width=10cm]{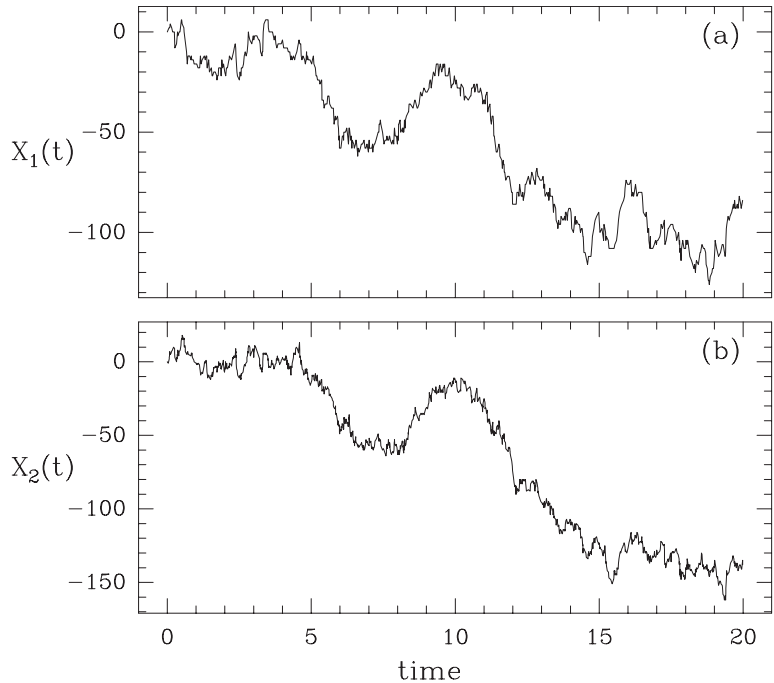}
\end{center}
\caption{
\label{figc2}
Numerical simulation of a 2D symmetric price model using Hawkes processes with parameters
$\alpha_{12} = 0.23$, $\alpha_{13} = 0.05$, $\beta = 0.11$ and $\mu = 0.015$.
(a) Sample path of $X_1(t)$  (b) Sample path of $X_2(t)$. The processes
appear more correlated then in Fig. \ref{figc1} because the asymptotic correlation
of large scale increments is $\rho = 0.65$.
}
\end{figure}

A realization of this process bivariate $(X_1,X_2)$ over $T=20$ hours is represented in Fig. \ref{figc1}
with $\alpha_{12} = 0.23$, $\alpha_{13} = 0.05$, $\beta = 0.11$ and $\mu = 0.015$.
(let us note that $\mu$, $\alpha_{12}$, $\alpha_{13}$ and $\beta$ are all expressed in the same unit, namely $\text{seconds}^{-1}$).
According to Eq. (\ref{correl}), the asymptotic correlation between the (large scale) increments of $X_1$ and $X_2$ is $\rho \simeq 0.15$.
Thus the two components are only moderately correlated (at a visual level, the graphs do not look alike).
Fig. \ref{figc2} shows the realization of the bivariate process $(X_1,X_2)$ in the case  $\alpha_{12} = 0.23$, $\alpha_{13} = 0.05$, $\beta = 0.11$ and $\mu = 0.015$.
Eq. (\ref{correl}) shows that the correlation between the two components is quite strong: $\rho \simeq 0.65$. This can be clearly seen on the two graphs which look quite much alike.

\begin{figure}[h]
\begin{center}
\includegraphics[width=10cm]{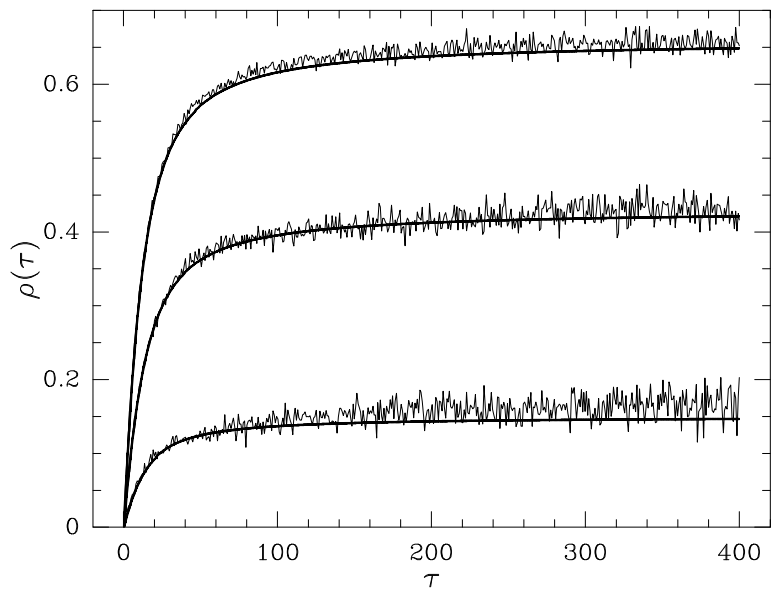}
\end{center}
\caption{
\label{figeppssimu}
Epps effect for simulation samples. The estimated correlation coefficients
are compared to the expected analytical curves (solid lines). From top top bottom
the cross correlation between the two asset returns increases from 0.15 to 0.65.
}
\end{figure}

In Fig. \ref{figeppssimu}(b),
the estimated and theoretical Epps effect  ({\it i.e.} the function $\rho(\tau) = C_{12}(\tau)/C_{11}(\tau)$)  are plotted for three sets of parameters, corresponding respectively to an asymptotic correlation coefficient of $0.15$, $0.40$ and $0.65$.

As for the univariate case,  one could perform maximum likelihood estimation of the parameters.  Bivariate formula for the likelihood are easy to obtained following the computations in Section \ref{est1d}.
In the bivariate model, the quantities of interest are not only the signature plots of both $X_1$ and $X_2$ but also the Epps effect between $X_1$ and $X_2$. The regression estimation now takes the form
\begin{align*}
\widehat \theta_{\text{reg}} = \text{Argmin}_{\theta}\big(& a_1|\widehat C_{11}(\tau) -  C_{11}(\tau) |^2 +a_2 |\widehat C_{22}(\tau) -  C_{22}(\tau) |^2 \\
&\;\;\;\;\;+a_{12} |\widehat C_{12}(\tau) -  C_{12}(\tau)) |^2\big),
\end{align*}
where  $C_{kl}(\tau) =  C_{kl}(\tau,0)$ is defined as in (\ref{cmatrix}) \textcolor{black}{and
\begin{equation}
   \widehat  C_{kl}(\tau)  = \frac{1}{T}\sum_{n=0}^{T/\tau} \Delta_\tau X_k (n\tau)\Delta_\tau X_l \big((n+1)\tau\big).
\end{equation}
The constants $a_1$,$a_2$ and $a_{12}$ are constant weights that are used to fix the relative minimization error of each term.}
Both estimators ($\widehat \theta_{\text{MLE}}$ and $\widehat \theta_{\text{reg}}$) lead to quite accurate results (with the same magnitude of errors in both cases).

\section{Diffusive (large scale) limit of the model}
\label{difflimit}
It is interesting to discuss the large scale limit of the processes obtained by our construction.
For that purpose, Let us define the following normalized processes for the general $N-$variate version of the model:
\begin{equation}
\{ X^{(T)}_i (t) \}_{i=1 \ldots N} =  \{ \frac{1}{\sqrt{T}} X_i(tT) \}_{i=1 \ldots N}\;\;\;\text{for}\;\;t \in [0,1].
\end{equation}
The question we would like to address concerns the existence and the properties of the \textcolor{black}{macroscopic limit $N$-variate process $\{X^{(\infty)}(t)\}_{i=1\ldots N}$ obtained by considering the large scale limit
$T \rightarrow \infty$.}
As far as the second order (correlation) properties of these processes are concerned, from Propositions \ref{prop1} and \ref{prop2}, one sees that in both the univariate and the bivariate case,
a diffusive limit exists. For instance, in the bivariate case, the limit process is
characterized by the following covariance matrix:
\begin{equation}
\label{Cmatrix}
{\mathbf C}^{(2)} = \left(
\begin{array}{ccc}
\displaystyle
\frac{2 A_{11}}{G_1}+ \frac{2 B_{11}}{G_2} & \frac{A_{12}+A_{21}}{G_1}+\frac{B_{12}+B_{21}}{G_2} \\
\frac{A_{12}+A_{21}}{G_1}+\frac{B_{12}+B_{21}}{G_2} & \frac{2 A_{22}}{G_1}+ \frac{2 B_{22}}{G_2}
\end{array}
\right),
\end{equation}
\textcolor{black}{By using limit theorem for semimartingales, it is actually possible to rigorously obtain a limiting process which is a multivariate Brownian motion with the appropriate covariance matrix. We describe and prove this results in details in a forthcoming paper}.

\section{Comparison to empirical data}
\label{data}
\subsection{Presentation of the data}
\label{dataset}
Let us recall that our goal is not to reproduce {\em perfectly} the
features observed from empirical data since on the one hand,  ultra high-frequency
tick-by-tick series can be defined in various way ({\it e.g.} mid prices, transaction prices,
ask or bid prices and so on), each of them having slightly different properties and representing
an arbitrary projection of the complex order book dynamics. On the other hand, it is
well know that market intraday fluctuations are not stationary and are characterized
by strong long term heterogeneities. For
computation and parsimony purposes, our model is somehow naive since the exponential
shape of the kernels $\varphi_{ij}$ \textcolor{black}{are somewhat arbitrary} and the chosen symmetries are arbitrary and not necessarily suitable in practice. We rather aim at providing a first-brick model that can be considered as
a tool to understand how the discrete nature of price variations at fine scales can be aggregated
at coarser scales and lead to cross correlated Brownian diffusion processes.
In that respect, comparisons to real data have to be interpreted rather at a qualitative level.

The data that have been used in this paper consist in tick-by-tick last traded price time series, with, for each trade,
the corresponding volume and a flag indicating whether the trade corresponds to a sell order or a buy order.
The prices are either Eurex Euro-Bund futures contracts or Eurex Euro-Bobl futures contracts which correspond respectively to long-term (8.5-10.5 years)
or medium-term (4.5 to 5.5 years) debt instrument issued by the Federal Republic of Germany.
%For a given date, only the most liquid (i.e., higer volume) contract is considered (most of the time it corresponds to the upfront month contract). The Euro-Bund is one of the most liquid instrument in the world (volume is greater than 300 million contracts in 2007).
Euro-Bund and Euro-Boble are well known to have highly correlated price variations. So they are good candidates to study the Epps effect.
Both open from (local-time) 8am to 10pm, but we shall restrict to the most liquid period : 8am to 5:15pm.
However, the liquidity (and the volatility) is highly seasonal during the day.
Our model does not account for such a seasonality, so we shall restrict the data to intraday periods \textcolor{black}{for which the underlying stationarity assumption is reasonable}. \textcolor{black}{Based on rough empirical considerations}, we select the time period 9am to 11am.
Moreover all computations have been made on last traded prices of buy orders only.
Choosing sell orders would not have change the results, however, taking into account in the same time-series both buy and sell orders would lead to a highly bouncing artefact that shall not be able to be captured by our modelling approach.

To summarize, the computations are made on two datasets :
\begin{itemize}
\item {\bf Dataset I} : 9am to 11pm from 11/01/2009 to 12/15/2009 (21 days) on the contract maturity 12/2009
\item {\bf Dataset II} : 9am to 11pm from 06/01/2009 to 08/01/2009 (41 days) on the contract maturities 06/2009 and 09/2009 (for each day the most liquid maturity is selected).
\end{itemize}

\subsection{Signature plots and Epps effect}
\label{realest}

\begin{figure}
\begin{center}
\includegraphics[width=10cm]{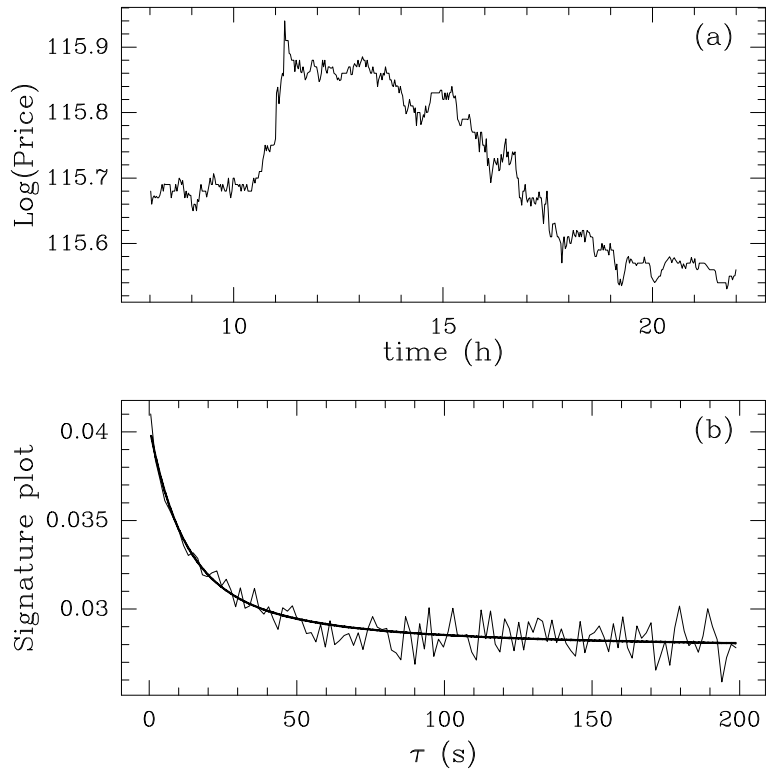}
\end{center}
\caption{
\label{figsplot1}
(a) Last traded (only buy market orders) prices path of the Euro-Bund contract on the  3rd of November 2009  (contract maturity 12/2009) from 8am to 10pm. (b) Associated mean daily signature plot (computed on the dataset I described at the end of Section \ref{dataset}).
The solid line represents the fit using the regression estimator $\theta_{reg}$ for the 1D Hawkes model (Section \ref{est1d}).
}
\end{figure}

Figure \ref{figsplot1}(a) shows the logarithm of the last traded price (only buy orders) of the Euro-Bund during a whole day 11/03/2009 (the contract maturity is 12/2009). The signature plot displayed in Fig. \ref{figsplot1}(b) has been computed using the dataset I described at the end of Section \ref{dataset}. Each day is considered to be an independent and identically distributed realization of the same process. In solid line we have superimposed
the signature plot obtained \textcolor{black}{when we fit the mean theoretical signature plot by a regression model in the univariate version of the model.}
\textcolor{black}{The estimates we obtain are $\widehat \mu = 0.016$, $\widehat \alpha = 0.024$ and $\widehat \beta = 0.11$.
We see that the curve associated with the model fits the data relatively well. Notice
that the MLE method also provides fairly good results (in that case, we obtain $\widehat \mu = 0.014$, $\widehat \alpha = 0.030$ and $\widehat \beta = 0.08$). However, as explained previously, one expects the MLE to be less stable with respect to ``noise" than the regression method on the mean signature plot and we empirically observe that its performance are worse than for the regression estimator. On other instances, they could lead to dramatically bad results and this is the reason why we discard the MLE estimator further on}.

\begin{figure}
\begin{center}
\includegraphics[width=10cm]{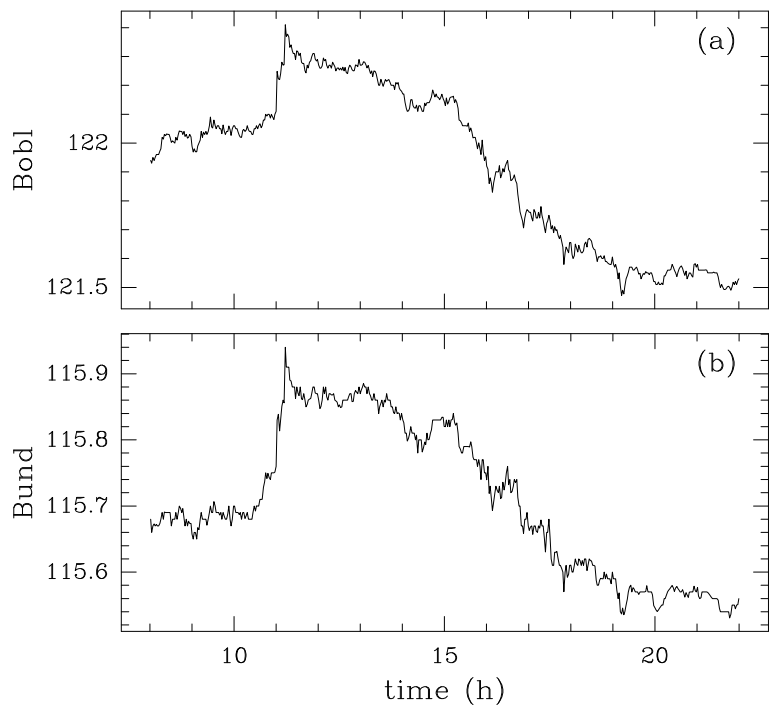}
\end{center}
\caption{
\label{figcharts}
(a)  Last traded (only buy orders) prices path of the Euro-Bobl contract on the  3rd of November 2009  (contract maturity 12/2009) from 8am to 10pm. (b) Same as (a) but for the Euro-Bund.
}
\end{figure}

\begin{figure}
\begin{center}
\includegraphics[width=10cm]{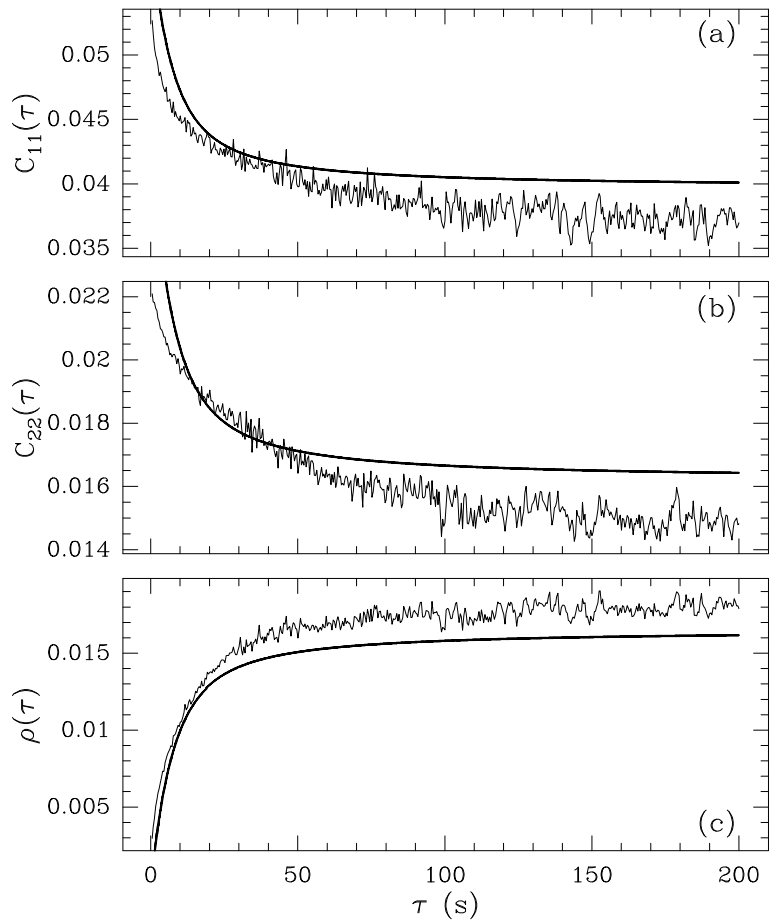}
\end{center}
\caption{
\label{fig2D} Self and cross correlation between Euro-Bund et Euro-Bobl returns
as a function of the time scale $\tau$. The computations have been made using the on the dataset II  described at the end of Section \ref{dataset}.
(a) Signature plot associated with the Euro-Bund  (b) Same as in (a)
for the Euro-Bobl. (c) Epps effect between Euro-Bund and Euro-Bobl as measured
by the covariance $\frac{C_{12}}{\tau}$ between the two asset returns
at different scales $\tau$. In (a), (b) and (c) the heavy lines represent a fit of the empirical curves according to the 2D Hawkes model (Proposition \ref{prop2}).
}
\end{figure}

Figure \ref{figcharts} shows the logarithm of the last traded price (only buy orders) of the Euro-Bund top) and Euro-Bobl (bottom) during a whole day 11/03/2009 (maturity 12/2009). One can directly observe the return correlations between
the two assets. The large scale correlation coefficient we find is close to $\rho = 0.77$.
In Figures \ref{fig2D}(a) and \ref{fig2D}(b) we plot the signature plots associated with the two assets while \ref{fig2D}(c) displays the estimated Epps effect as measured by the correlation coefficient \textcolor{black}{$\widehat C_{12}(\tau)$} at different
scales. The computations were made using the dataset II described at the end of Section \ref{dataset}. The solid lines represent the fits according to the regression method of the bivariate model. \textcolor{black}{One can see that although significant discrepencies between empirical and fitted data are observed,
given the simplicity of the model, one can consider that it captures fairly well
both variance and covariance features of assets from small to large time scales simultaneously}.

\begin{figure}
\begin{center}
\includegraphics[width=10cm]{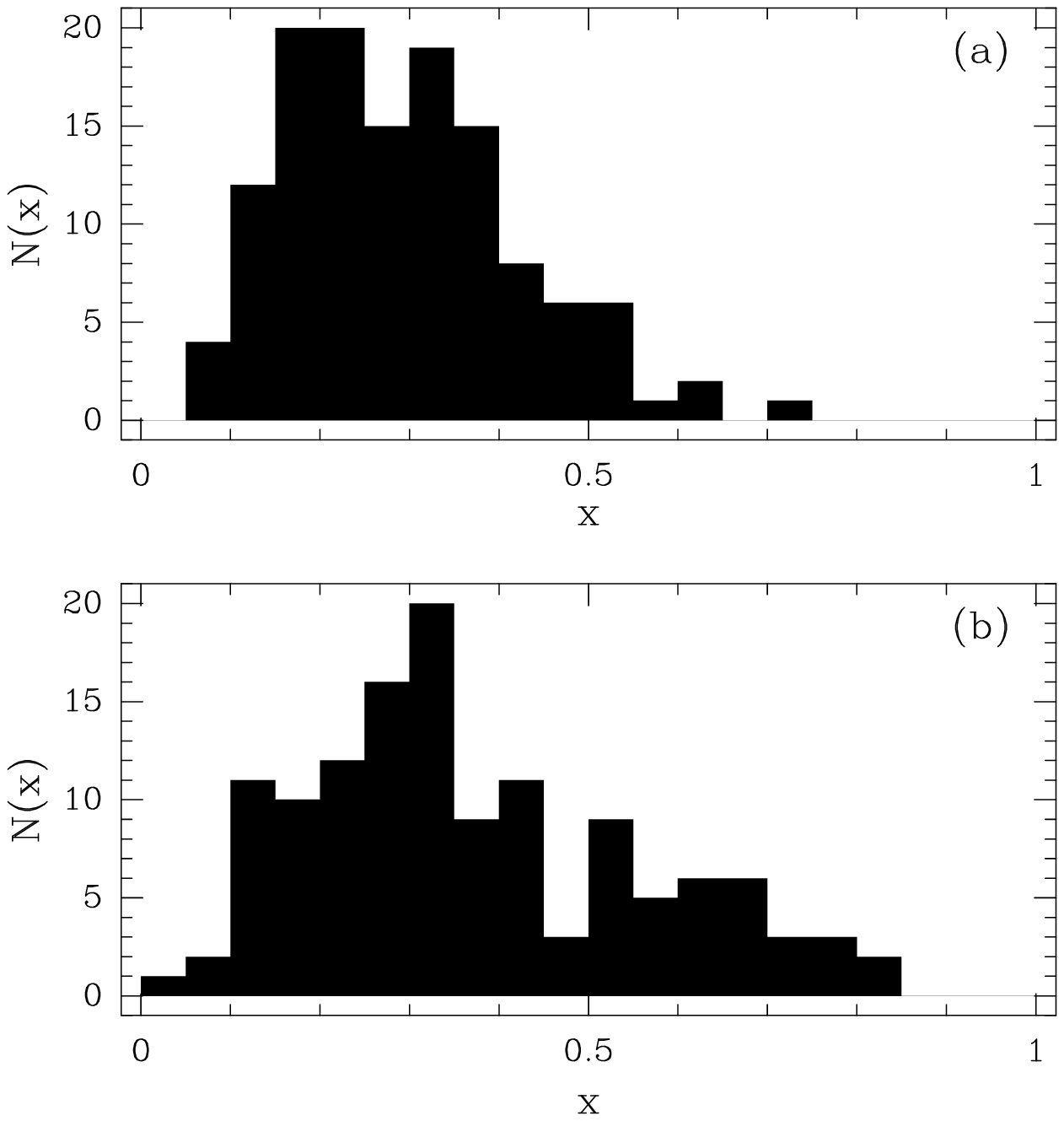}
\end{center}
\caption{
\label{figh}
(a) Histogram of values of $x$ (as defined in Eq. \eqref{defx}) for the Bund front month contract from 5/14/2009 to 12/31/2009. (c) Same plot
for the Bobl over the same period. The mean values for the two assets are respectively $x = 0.29$ and $x = 0.36$.
}
\end{figure}

\section{Conclusion and prospects}
\label{conclusion}

We have proposed in this paper a simple bivariate tick-by-tick price model
based on Hawkes (self and mutually exciting) point processes. We have shown that closed
form expressions can be obtained for its second order properties at all time scales.
This allows one to recover major high frequency stylized facts, namely the
signature plot behavior and the Epps effect. We have shown that the model is
easy to simulate and can be estimated with a MLE method or using a moment method.
When compared to real data, we have seen that the 2D model hardly accounts
for the exact behavior of signature plots and correlation functions.
However, as emphasized in section \ref{data}, the approach introduced in this
paper has to be considered as a simple framework that allows one to address
issues related to the relationship between fine and coarse scale properties of
market dynamics. It can also be used as a simple tool in order to investigate
 intraday market features using few parameters that are easy to interpret.
For instance, according to the univariate version of the model, the variance associated
with the microstructure is
$$\Lambda = \frac{2\mu}{1-||\varphi||_1}$$ while the large scale diffuse
volatility is
$$\Lambda \kappa^2 = \frac{2\mu}{1-||\varphi||_1}\frac{1}{(1+||\varphi||_1)^2}.$$
One sees that the mean reversion (as measured by $0 < ||\varphi||_1 < 1$) is softened at large scale
thanks to the diffusion. however, the influence of $\varphi$ does not completely disappear. \textcolor{black}{From the above tow expressions of the microscopic and the macroscopic variance,} the large scale
effect of mean reversion can be quantified by the function:
\begin{equation}
\label{defx}
    ||\varphi||_1 = x \in [0,1) \rightarrow \frac{1}{(1-x)(1+x)^2}
\end{equation}
which has a minimum around $x = 1/3$.
In Fig. \ref{figh} we plot the histograms of estimated values of $||\varphi||_1$
during a period of 6 months for Bund and Bobl front month contracts (from 05/14/2009 to 12/31/2009). It is striking to observe
that both distributions are peaked (with rather large deviations however) around $x = 1/3$.
This suggests that the market chooses the strength of microstructural mean reversion in
order to minimize its long term volatility.

In a future work, we will address the questions related to lead-lag effects
that can be easily accounted, for each time scale, within our model.
It will also be interesting to consider non parametric estimates of the
kernel shapes ($\varphi_{ij}$) along the line of the work of \cite{Bouret} and
to consider various questions related to the well known non stationarity and long-range correlations
of the volatility. Let us stress that it is also tempting to bridge the parametric approach advocated in this paper to the recent work of Joulin et al. \cite{Jou08} where the authors studied and quantified the effects of exogeneous news
with respect to the endogeneous noise on the jumps of stock prices.

\section*{Appendix 1: Signature plot in the univariate case}
In the univariate case the shape of the signature plot can be directly computed for an exponential kernel $\varphi$ as given by Eq. \eqref{expokernel}.
Indeed, if one defines
\begin{equation}
   \bar \Lambda  =  \mathbb{E}\left[\frac{dN_{1}}{dt} \right] = \mathbb{E}\left[\frac{dN_{2}}{dt} \right]
\end{equation}
it directly results from Eqs \eqref{defl1} and \eqref{defl2} that
\begin{equation}
   \bar \Lambda = \frac{\mu}{1-||\varphi||_1} = \frac{\mu \beta}{\beta-\alpha} \; .
\end{equation}
From the definition of the signature plot and by symmetry between the processes $N_1(t)$ and $N_2(t)$,
we have, \textcolor{black}{for $\tau >0$}:
\begin{equation}
C(\tau) = \frac{2}{\tau} \left( \mathbb{E}\left[ N_1^2(\tau) \right] - \mathbb{E}\left[ N_1(\tau) N_2(\tau) \right] \right)
\end{equation}
Let $M_{ij}(t)$ denotes the continuous part of the covariance of $dN_i(t)$ and $dN_j(t)$ and $M(t)= M_{11}(t)-M_{12}(t)$. From the definition of
$N_1(t)$ and $N_2(t)$, we then have:
\begin{eqnarray*}
   \mathbb{E} \left[ N_1(\tau)^2 \right] & = & \bar \Lambda t + \bar\Lambda^2 \tau^2 + \int_0^{\tau} \int_0^{\tau} M_{11}(u-v) du dv \\
   \mathbb{E}\left[ N_1(\tau) N_2(\tau) \right] & = & \bar \Lambda^2 \tau^2 + \int_0^{\tau} \int_0^{\tau} M_{12}(u-v) du dv
\end{eqnarray*}
and
\begin{equation}
\label{sp1}
C(\tau) = 2 \left( \bar \Lambda + \tau^{-1} \int_0^{\tau} \int_0^{\tau} M_{12}(u-v) du dv \right).
\end{equation}
In order to compute $M_{ij}(\tau)$ and $M(\tau)$, one can first estimate the conditional mean with respect to ${\mathcal F}_{t+\tau}$ and then perform unconditional averages. Using this trick, from the definition of $\lambda_1(t)$ and $\lambda_2(t)$, one finds:
\begin{eqnarray*}
  M_{11}(\tau) & = & \int_{-\infty}^{\tau} \varphi(\tau-u) M_{12}(u) \; du \\
  M_{12}(\tau) & = & \bar \Lambda \varphi(\tau ) + \int_{-\infty}^{\tau } \varphi(\tau -u) M_{11}(u)du
\end{eqnarray*}
and consequently, $M(\tau )$ satisfies:
\begin{equation}
    M(\tau)  = -\varphi(\tau) \bar \Lambda + \int_{-\infty}^{\tau} \varphi(\tau -u) M(u) \; du \;.
\end{equation}
If one seeks for an exponential solution $M(x) = a e^{-bx}$ to this equation, one finds:
\begin{eqnarray*}
     b & = & -(\alpha+\beta) \\
     a & = & -\frac{\bar \Lambda \alpha (\alpha+2\beta)}{2(\alpha+\beta)}
\end{eqnarray*}
and finally, from \eqref{sp1},
\begin{equation}
\textcolor{black}{C(\tau) = 2\bar \Lambda \left[ \frac{1}{(1+\frac{\alpha}{\beta})^2}+ \left(1-\frac{1}{(1+\frac{\alpha}{\beta})^2}\right)\frac{1-e^{-(\alpha+\beta)\tau}}{(\alpha+\beta)\tau}\right]}.
\end{equation}

\section*{Appendix 2: Correlation function in the multivariate case}
Let us define ${\mathbf M}(\tau)$ as the (continuous part of the) covariance matrix
of individual point densities and  ${\bf \Lambda}$ the mean intensity vector:
\begin{eqnarray*}
     \Lambda_{i} & = & \mathbb{E}\left[\frac{dN_{i}}{dt} \right] \\
     M_{ij}(u) & = & \mbox{Cov} \left[\frac{dN_{i}(t_0)}{dt},\frac{dN_{j}(t_0+u)}{dt}\right]
\end{eqnarray*}

It is straightforward to show that $\mathbf{C}$ can be obtained in terms of $\mathbf{M}$ as:
\begin{equation}
\label{covX}
    C_{\rho \nu}(\tau,t) = \int_0^\tau du \int_{t}^{t+\tau} dv \; K_{\rho \nu}(v-u) \; ,
\end{equation}
where the matrix $\mathbf{K}$ is
\begin{eqnarray}
\label{defK}
\nonumber
  & K_{11}(z) = & (\Lambda_{1}+\Lambda_{2})\delta(z) + J_{11}(z)  \\
\nonumber
 & K_{12}(z)  = & J_{12}(z)  \\
&  K_{21}(z)  = & J_{21}(z) \\
\nonumber
 & K_{22}(z)  = & (\Lambda_{3}+\Lambda_{4})\delta(z) + J_{22}(z)
\end{eqnarray}
with $\mathbf{J}$ defined as:
\begin{eqnarray}
\label{defJ}
\nonumber
  & J_{11}(z) = & M_{11}(z)+M_{22}(z)-M_{12}(z)-M_{21}(z)  \\
\nonumber
 & J_{12}(z)  = & M_{13}(z)+M_{24}(z)-M_{23}(z)-M_{14}(z)  \\
&  J_{21}(z)  = & M_{31}(z)+M_{42}(z)-M_{41}(z)-M_{32}(z) \\
\nonumber
 & J_{22}(z)  = & M_{33}(z)+M_{44}(z)-M_{34}(z)-M_{43}
\end{eqnarray}

Along the same line as in \cite{Hawkes}, one can show that the matrix ${\mathbf M}$ satisfies
the  following integral equation:
\begin{equation}
\label{iterm}
   \mathbf{M}(u) = \mathbf{\Phi}(u) \; \mbox{diag} \left({\mathbf \Lambda}\right) +\int_{-\infty}^{u} {\mathbf \Phi}(u-v) {\mathbf M}(v) \; dv \; \; \mbox{for} \; \; u > 0.
\end{equation}
The values for $u<0$ can be obtained thanks to the equality
$M_{ij}(-v) = M_{ji}(v)$ that is a direct consequence of the definition of $\mathbf{M}$.

Let us denote by $\tilde{F}(s)$ the unilateral Laplace
transform of $F(u)$.
By taking the Laplace transform of Eq. \eqref{iterm},
one gets the following set of coupled linear equation
of $\tilde{{\mathbf M}}(s)$:
\begin{equation}
\label{laplace}
   {\tilde M_{ij}}(s) = \Lambda_j {\tilde \varphi}_{ij}(s)+\sum_{k} {\tilde \varphi}_{ik}(s) \left( {\tilde M}_{kj}(s)+{\tilde M}_{jk}(\beta_{ik})  \right).
\end{equation}
By using Eqs. \eqref{defJ},
it follows that the Laplace transform of $J_{\rho \nu}$ satisfies the following linear system:
\begin{eqnarray*}
  {\tilde J}_{11}(s) & = & -{\tilde \varphi}_{12}(s){\tilde J}_{11}(s)
   +{\tilde \varphi}_{13}(s) {\tilde J}_{21}(s) -{\tilde \varphi}_{12}(s)\left[ \Lambda_1+\Lambda_2+{\tilde J}_{11}(\beta_{12})\right] \\ & &+ {\tilde \varphi}_{13}(s) {\tilde J}_{12}(\beta_{13}) \\
  {\tilde J}_{12}(s) & = & -{\tilde \varphi}_{12}(s){\tilde J}_{12}(s)+{\tilde \varphi}_{13}(s){\tilde J}_{22}(s)
   +{\tilde \varphi}_{13}(s)\left[\Lambda_{3}+\Lambda_4+{\tilde J}_{22}(\beta_{13}) \right] \\ & &
   -{\tilde \varphi}_{12}(s) {\tilde J}_{21}(\beta_{12}) \\
  {\tilde J}_{21}(s) & = & -{\tilde \varphi}_{34}(s){\tilde J}_{21}(s)+{\tilde \varphi}_{31}(s){\tilde J}_{11}(s)
   +{\tilde \varphi}_{31}(s)\left[\Lambda_{1}+\Lambda_2+{\tilde J}_{11}(\beta_{31}) \right] \\ &&-{\tilde \varphi}_{34}(s) {\tilde J}_{12}(\beta_{34}) \\
  {\tilde J}_{22}(s) & = & -{\tilde \varphi}_{34}(s){\tilde J}_{22}(s)+{\tilde \varphi}_{31}(s) {\tilde J}_{12}(s) -{\tilde \varphi}_{34}(s)\left[\Lambda_3+\Lambda_4+{\tilde J}_{22}(\beta_{34}) \right] \\
  & &  +{\tilde \varphi}_{31}(s) {\tilde J}_{21}(\beta_{31}).
\end{eqnarray*}
The solution to this system reads:
\begin{equation}
\label{cc}
{\tilde J}(s) =
\left( \begin {array}{c} {\frac {{  v_1}+{  v_1}\,{  \varphi_{34}}+{  \varphi_{13}}\,{  v_3}}{1+{  \varphi_{12}}+{  \varphi_{34}}+{  \varphi_{34}
}\,{  \varphi_{12}}-{  \varphi_{31}}\,{  \varphi_{13}}}}\\\noalign{\medskip}{
\frac {{  v_2}+{  v_2}\,{  \varphi_{34}}+{  \varphi_{13}}\,{  v_4}}
{1+{  \varphi_{12}}+{  \varphi_{34}}+{  \varphi_{34}}\,{  \varphi_{12}}-{
\varphi_{31}}\,{  \varphi_{13}}}}\\\noalign{\medskip}{\frac {{  \varphi_{31}}\,{
  v_1}+{  v_3}+{  v_3}\,{  \varphi_{12}}}{1+{  \varphi_{12}}+{
\varphi_{34}}+{  \varphi_{34}}\,{  \varphi_{12}}-{  \varphi_{31}}\,{  \varphi_{13}}}}
\\\noalign{\medskip}{\frac {{  \varphi_{31}}\,{  v_2}+{  v_4}+{
v_4}\,{  \varphi_{12}}}{1+{  \varphi_{12}}+{  \varphi_{34}}+{  \varphi_{34}}\,{
  \varphi_{12}}-{  \varphi_{31}}\,{  \varphi_{13}}}}\end {array} \right),
\end{equation}
where we have denoted by ${\mathbf v}$ the vector
\begin{equation}
\label{vv}
{\mathbf v} =
\left( \begin {array}{c} -{\tilde \varphi}_{12}(s)\left( \Lambda_1+\Lambda_2+{\tilde J}_{11}(\beta_{12})\right) + {\tilde \varphi}_{13}(s) {\tilde J}_{12}(\beta_{13})
\\ {\tilde \varphi}_{13}(s)\left(\Lambda_{3}+\Lambda_4+{\tilde J}_{22}(\beta_{13}) \right) -{\tilde \varphi}_{12}(s) {\tilde J}_{21}(\beta_{12})  \\
{\tilde \varphi}_{31}(s)\left(\Lambda_{1}+\Lambda_2+{\tilde J}_{11}(\beta_{31}) \right) -{\tilde \varphi}_{34}(s) {\tilde J}_{12}(\beta_{34}) \\
-{\tilde \varphi}_{34}(s)\left(\Lambda_3+\Lambda_4+{\tilde J}_{22}(\beta_{34}) \right) +{\tilde \varphi}_{31}(s) {\tilde J}_{21}(\beta_{31})
\end {array} \right).
\end{equation}
This vector can be computed if by evaluating ${\mathbf J}(s)$ at $s = \beta_{12},\ldots,\beta_{34}$.
If one now considers that the Laplace transform of ${\tilde \varphi}_{ij}(x)$
is
\begin{equation}
   {\tilde \varphi}_{ij}(s) = \frac{\alpha_{ij}}{s+\beta_{ij}} \; ,
\end{equation}
it is possible to compute
the expression of each component $J_{\alpha \beta}$
and, by the inverse Laplace transform, to obtain the
correlation matrix $C_{\alpha \beta}(t,\tau)$.
Accounting for the symmetries in Eq. \eqref{phimatrix},
in Eq. \eqref{vv}, we now have 4 constants to determine: $Q_1 = {\tilde J_{11}}(\beta)$, $Q_2 = {\tilde J_{12}}(\beta)$,
$Q_3 = {\tilde J_{21}}(\beta)$ and $Q_4 = {\tilde J_{22}}(\beta)$.
If we solve the linear system obtained by evaluating \eqref{cc} in $s=\beta$, we obtain,
{\tiny
\begin{eqnarray*}
Q_1 & = &  -\frac{\Lambda_1\left(\alpha_{31} \alpha_{13}(1+\alpha_{34}+\beta+\alpha_{12})
-3 \beta \alpha_{34} \alpha_{12} -\beta \alpha_{12}^2-\alpha_{34}^2 \alpha_{12}
-2 \alpha_{12} \beta^2-\alpha_{12}^2 \alpha_{34}\right)+\beta \Lambda_3 \alpha_{13}^2}{\alpha_{13} \alpha_{31}(2 \beta +\alpha_{12}+\alpha_{34}) -3 \beta^2 \alpha_{12}-\alpha_{12}^2 \alpha_{34}-\alpha_{34}^2 \alpha_{12}
-3 \beta^2 \alpha_{34}-2 \beta^3-4 \beta \alpha_{34} \alpha_{12}-\beta \alpha_{34}^2-\alpha_{12}^2 \beta} \\
Q_2 & = &-\frac{\Lambda_3(2 \beta^2\alpha_{13}-\alpha_{31}\alpha_{13}^2 +2 \alpha_{12} \beta \alpha_{13}+\alpha_{34} \alpha_{12}\alpha_{13}+\alpha_{34} \beta \alpha_{13})
+\Lambda_1(\alpha_{13} \alpha_{31}^2
-\alpha_{12} \beta \alpha_{31}
-\alpha_{31} \alpha_{34} \alpha_{12})}
{\alpha_{13} \alpha_{31}(2 \beta +\alpha_{12}+\alpha_{34}) -3 \beta^2 \alpha_{12}-\alpha_{12}^2 \alpha_{34}-\alpha_{34}^2 \alpha_{12}
-3 \beta^2 \alpha_{34}-2 \beta^3-4 \beta \alpha_{34} \alpha_{12}-\beta \alpha_{34}^2-\alpha_{12}^2 \beta} \\
Q_3 & = &-\frac{\Lambda_1(2 \beta^2\alpha_{31}-\alpha_{31}\alpha_{13}^2 +2 \alpha_{34} \beta \alpha_{31}+\alpha_{34} \alpha_{12}\alpha_{31}+\alpha_{12} \beta \alpha_{31})
+\Lambda_3(\alpha_{13} \alpha_{31}^2
-\alpha_{34} \beta \alpha_{13}
-\alpha_{13} \alpha_{34} \alpha_{12})}
{\alpha_{13} \alpha_{31}(2 \beta +\alpha_{12}+\alpha_{34}) -3 \beta^2 \alpha_{12}-\alpha_{12}^2 \alpha_{34}-\alpha_{34}^2 \alpha_{12}
-3 \beta^2 \alpha_{34}-2 \beta^3-4 \beta \alpha_{34} \alpha_{12}-\beta \alpha_{34}^2-\alpha_{12}^2 \beta} \\
Q_1 & = &  -\frac{\Lambda_3\left(\alpha_{31} \alpha_{13}(1+\alpha_{12}+\beta+\alpha_{34})
-3 \beta \alpha_{34} \alpha_{12} -\beta \alpha_{34}^2-\alpha_{12}^2 \alpha_{34}
-2 \alpha_{34} \beta^2-\alpha_{34}^2 \alpha_{12}\right)+\beta \Lambda_1 \alpha_{34}^2}{\alpha_{13} \alpha_{31}(2 \beta +\alpha_{12}+\alpha_{34}) -3 \beta^2 \alpha_{12}-\alpha_{12}^2 \alpha_{34}-\alpha_{34}^2 \alpha_{12}
-3 \beta^2 \alpha_{34}-2 \beta^3-4 \beta \alpha_{34} \alpha_{12}-\beta \alpha_{34}^2-\alpha_{12}^2 \beta}.
\end{eqnarray*}
}
where the expressions for the $\Lambda_i$ are provided in Eq. \eqref{lambda}.
One can now inverse the Laplace transforms in Eq. \eqref{cc} and
and, from Eq. \eqref{defK} the expressions of the functions $K_{\rho \nu}(t)$. This leads to
\begin{eqnarray*}
   K_{11}(t) & = & 2 \Lambda_1 \delta(t) + A_{11} e^{-G_1 t} + B_{11} e^{-G_2 t}  \\
   K_{12}(t) & = & A_{12} e^{-G_1 t}  +B_{12}  e^{-G_2 t}  \\
   K_{21}(t) & = & A_{21} e^{-G_1 t}  +B_{21}  e^{-G_2 t} \\
   K_{22}(t) & = & 2 \Lambda_3 \delta(t)+ A_{22} e^{-G_1 t}  +B_{22}  e^{-G_2t}
\end{eqnarray*}
where
\begin{eqnarray*}
   G_1 & = & Y+Z \\
   G_2 & = & Y-Z \\
   Y  & = &  \beta + \frac{1}{2}(\alpha_{12}+\alpha_{34}) \\
   Z  & = &  \frac{1}{2}\sqrt{(\alpha_{12}-\alpha_{34})^2+4 \alpha_{13} \alpha_{31}}
\end{eqnarray*}
and the expressions of constants $A_{ij}$, $B_{ij}$ read:
\begin{eqnarray*}
A_{11} & =& -\big(Q_1(2\alpha_{13}\alpha_{31}+\alpha_{12}^2+2 Z \alpha_{12}-\alpha_{34} \alpha_{12})\\ & &-Q_2
(\alpha_{12} \alpha_{13}+2 Z \alpha_{13}+\alpha_{34} \alpha_{13}) \\
& & +\Lambda_1 (4 \alpha_{13} \alpha_{31}+2 \alpha_{12}^2+4 Z \alpha_{12}-2 \alpha_{34} \alpha_{12})\big)/(4Z) \\
B_{11} & = &  - \big(Q_1 (\alpha_{34} \alpha_{12}-2 \alpha_{13} \alpha_{31}-\alpha_{12}^2+2 Z \alpha_{12}) \\ &&+
Q_2 (\alpha_{13} \alpha_{12}-2 Z \alpha_{13}+\alpha_{34} \alpha_{13})+ \\
& & +\Lambda_1
(2 \alpha_{34} \alpha_{12}+4 Z \alpha_{12}-2 \alpha_{12}^2-4 \alpha_{13} \alpha_{31})\big)/(4Z) \\
A_{12} & = &- \big(Q_3 (2 \alpha_{13} \alpha_{31}+2 Z \alpha_{12}+\alpha_{12}^2-\alpha_{34} \alpha_{12})
\\ &&-Q_4 (\alpha_{13} \alpha_{12}+2 Z \alpha_{13}+\alpha_{34} \alpha_{13})\\
& & -\Lambda_3
(4 Z \alpha_{13}+2 \alpha_{34} \alpha_{13}+2 \alpha_{12} \alpha_{13})\big)/(4Z) \\
B_{12} &= &-\big(Q_3 (\alpha_{34} \alpha_{12}-2 \alpha_{13} \alpha_{31}+2 Z \alpha_{12}-\alpha_{12}^2)\\ &&+Q_4 (\alpha_{34} \alpha_{13}+\alpha_{12} \alpha_{13}-2 Z \alpha_{13}) \\
& & +\Lambda_3 (2 \alpha_{13} \alpha_{12}-4 Z \alpha_{13}+2 \alpha_{13} \alpha_{34})\big)/(4Z) \\
A_{21} & = & \big(Q_1 (\alpha_{12} \alpha_{31}+2 Z \alpha_{31}+\alpha_{34} \alpha_{31})\\ &&+Q_2 (\alpha_{12} \alpha_{34}-\alpha_{34}^2-2 \alpha_{13} \alpha_{31}-2 Z \alpha_{34}) \\
& & +\Lambda_1 (2 \alpha_{12} \alpha_{31}+2 \alpha_{34} \alpha_{31}+4 Z \alpha_{31})\big)/(4Z) \\
B_{21} &=& \big(Q_2 (\alpha_{34}^2-\alpha_{12} \alpha_{34}+2 \alpha_{13} \alpha_{31}-2 Z \alpha_{34}) \\ && -Q_1 (\alpha_{12} \alpha_{31}-\alpha_{34} \alpha_{31}+2 Z \alpha_{31}) \\
& & +\Lambda_1
(4 Z \alpha_{31}-2 \alpha_{12} \alpha_{31}-2 \alpha_{34} \alpha_{31})\big)/(4Z) \\
A_{22} &= & \big(Q_4 (\alpha_{12} \alpha_{34}-2 \alpha_{13} \alpha_{31}-2 Z \alpha_{34}-\alpha_{34}^2) \\ &&+Q_3 (\alpha_{31} \alpha_{34}+2 Z \alpha_{31}+\alpha_{31} \alpha_{12}) \\
& & +\Lambda_3
(2 \alpha_{12} \alpha_{34}-2 \alpha_{34}^2-4 \alpha_{13} \alpha_{31}-4 Z \alpha_{34})\big)/(4Z) \\
B_{22} &= & \big(Q_4 (2 \alpha_{13} \alpha_{31}-\alpha_{12} \alpha_{34}+\alpha_{34}^2-2 Z \alpha_{34})\\ &&+Q_3 (2 Z \alpha_{31}-\alpha_{12} \alpha_{31}-\alpha_{34} \alpha_{31}) \\
& & +\Lambda_3
(4 \alpha_{13} \alpha_{31}-4 Z \alpha_{34}+2 \alpha_{34}^2-2 \alpha_{12} \alpha_{34})\big)/(4Z).
\end{eqnarray*}
By performing the double integral \eqref{covX}, one finally obtains:
{\footnotesize
\begin{eqnarray*}
\frac{C_{11}(\tau)}{\tau} & = & \frac{2 A_{11}e^{-G_1\tau}}{G_1^2 \tau}+
\frac{2 B_{11}e^{-G_2\tau}}{G_2^2\tau}+2 \Lambda_1
+\frac{2 A_{11}}{G_1}+\frac{2 B_{11}}{G_2}
-\frac{2 A_{11}}{G_1^2\tau}-\frac{2 B_{11}}{G_2^2\tau} \\
\frac{C_{21}(\tau)}{\tau} = \frac{C_{12}(\tau)}{\tau} & = & (A_{21}+A_{12})(\frac{e^{-G_1\tau}-1}{G_1^2 \tau}
+\frac{1}{G_1})
+(B_{21}+B_{12})(\frac{e^{-G_2\tau}-1}{G_2^2\tau}+\frac{1}{G_2})\\
\frac{C_{22}(\tau)}{\tau} & = & \frac{2 A_{22}e^{-G_1\tau}}{G_1^2 \tau}+
\frac{2 B_{22}e^{-G_2\tau}}{G_2^2\tau}+2 \Lambda_3
+\frac{2A_{22}}{G_1}+\frac{2B_{22}}{G_2}
-\frac{2 A_{22}}{G_1^2\tau}-\frac{2 B_{22}}{G_2^2\tau}.
\end{eqnarray*}
}

\textcolor{black}{{\small \noindent {\bf Acknowledgements}. We thank Mathieu Rosenbaum for helpful discussions. M.H. wishes to thank Marek Musiela and the Electronic Trading Group in the FIRST team of BNP-Paribas for stimulating discussions on modelling dependence structures using point processes.}}
\bibliographystyle{plain}
% \bibliography{BDHM1}

\end{document}